\documentstyle[twoside,fleqn,espcrc2]{article}

\newcommand{\AmS}{{\protect\the\textfont2
  A\kern-.1667em\lower.5ex\hbox{M}\kern-.125emS}}
\def\lsim{\mathrel{\raise.3ex\hbox{$<$\kern-.75em\lower1ex\hbox{$\sim$}}}}
\def\gsim{\mathrel{\raise.3ex\hbox{$>$\kern-.75em\lower1ex\hbox{$\sim$}}}}
\def\ifmath#1{\relax\ifmmode #1\else $#1$\fi}
\def\ls#1{\ifmath{_{\lower1.5pt\hbox{$\scriptstyle #1$}}}}
\def\half{\ifmath{{\textstyle{1 \over 2}}}}

\def\tanb{\tan\beta}
\def\hl{h^0}
\def\ha{A^0}
\def\hh{H^0}
\def\hpm{H^\pm}
\def\mha{m_{\ha}}
\def\mhl{m_{\hl}}

\def\mz{m_Z}

\def\mt{m_t}

\def\msusy{M_{\rm SUSY}}
\def\mzz{m_Z^2}
\def\mww{m_W^2}

\def\mpl{M_{\rm P}}

\def\mx{M_{\rm X}}
\def\snu{\widetilde\nu}
\def\ls#1{\ifmath{_{\lower1.5pt\hbox{$\scriptstyle #1$}}}}
\def\msusy{M_{\rm SUSY}}
\def\opcit{{\it op.~cit.}}

\def\wh{\widehat}
\def\wt{\widetilde}
\def\pri{^{\prime}}

\let\ds=\displaystyle
\let\us=\underline
\newenvironment{Eqnarray}%
         {\arraycolsep 0.14em\begin{eqnarray}}{\end{eqnarray}}
\def\beq{\begin{equation}}
\def\eeq{\end{equation}}
\def\beqa{\begin{Eqnarray}}
\def\eeqa{\end{Eqnarray}}

\begin{document}

\vbox{  \large
\begin{flushright}
SCIPP 97/27 \\
hep--ph/9709450\\
\end{flushright}
\vskip1cm
\begin{center}
{\LARGE\bf
The Status of the Minimal Supersymmetric Standard Model and Beyond}\\[1pc]

{\bf Howard E. Haber}\\[2.mm] {\it Santa Cruz Institute for Particle
Physics,  \\ University of California, Santa Cruz, CA 95064}\\

\vskip2cm
\thispagestyle{empty}

{\bf Abstract}\\[1pc]

\begin{minipage}{15cm}
The minimal supersymmetric extension of the Standard Model (MSSM) is
reviewed.  In the most general framework with minimal field content and
R-parity conservation, the MSSM is a 124-parameter model (henceforth
called MSSM-124).  An acceptable phenomenology occurs
only at exceptional points (and small perturbations
around these points) of MSSM-124 parameter space.
Among the topics addressed in this review are: gauge coupling
unification, precision electroweak data, phenomenology of the MSSM Higgs
sector, and supersymmetry searches at present and future colliders.
The implications of approaches beyond the MSSM are briefly addressed.
\end{minipage}  \\
\vskip2cm
Invited Talk at the \\
5th International Conference on Supersymmetries in Physics (SUSY 97)
\\
27--31 May 1997, University of Pennsylvania, Philadelphia, PA USA \\
\end{center}
}
\vfill
\clearpage

\setcounter{page}{1}

\title{The Status of the Minimal Supersymmetric Standard Model and
Beyond}%

\author{Howard E. Haber\address{Santa Cruz Institute for Particle
        Physics \\
        University of California, Santa Cruz, CA 95064 USA}}


\begin{abstract}
The minimal supersymmetric extension of the Standard Model (MSSM) is
reviewed.  In the most general framework with minimal field content and
R-parity conservation, the MSSM is a 124-parameter model (henceforth
called MSSM-124).  An acceptable phenomenology occurs
only at exceptional points (and small perturbations
around these points) of MSSM-124 parameter space.
Among the topics addressed in this review are: gauge coupling
unification, precision electroweak data, phenomenology of the MSSM Higgs
sector, and supersymmetry searches at present and future colliders.
The implications of approaches beyond the MSSM are briefly addressed.
\end{abstract}

\maketitle

\section{UNIFICATION:~PAST~AND~PRESENT}

During the 1980s, a series of meetings were held called the $n$th
Workshop on Grand Unification (nWOGU).  By the end of the 1970s, the
electroweak model of Glashow, Weinberg and Salam had been confirmed
by the experimental observations of the SU(2)$\times$U(1) structure
of the weak neutral current.  Models of grand unified theories (GUTs) of
the strong and electroweak forces were being
studied intensely, with some indications for the minimal SU(5) unification
model.    In particular, the measurement of the weak mixing angle,
$\sin^2\theta_W$, seemed to in agreement with the SU(5) prediction.
At the 4th WOGU held in 1983 at the
University of Pennsylvania, Marciano \cite{4wogu} reported
the SU(5) prediction of:
\beq \label{e1}
\sin^2\theta_W(m_W)=0.214^{+0.004}_{-0.003}\quad [\rm{SU(5)~GUT]}\,,
\eeq
which was in excellent agreement with the experimental results based on
deep-inelastic $\nu N$ scattering and polarized electron deuteron
($ed$) scattering asymmetry measurements (including ${\cal O}(\alpha)$
radiative corrections):
\beqa \label{e2}
\sin^2\theta_W(m_W) = 0.215\pm 0.014 &&\quad [\nu N~{\rm data}]\,,
\nonumber \\
\sin^2\theta_W(m_W) = 0.216\pm 0.020 &&\quad [ed~{\rm data}]\,.
\eeqa

The apparent success of the SU(5) model encouraged theorists to take
seriously its predictions for proton decay.  There were already a couple
of candidates from existing proton decay experiments which had generated
some interest.  New proton decay
experiments with increased sensitivity and better resolution
were beginning to take data.  There were high
expectations that proton decay would soon be observed in the new
experiments.  The organizers of WOGU optimistically anticipated the
announcement of the discovery of proton decay during the early 1980s.
More detailed measurements of proton decay branching ratios were
expected later in the decade and would provide the necessary evidence to
either confirm the simplest SU(5) grand unified model or to distinguish
among more complicated GUT scenarios.

In 1983, although the success of the gauge coupling unification
prediction apparently provided strong evidence
for the GUT approach, no evidence for proton decay was forthcoming.
The IMB Collaboration reported $\Gamma^{-1}(p\to e^+\pi^0)> 6\times
10^{31}$~years, and Marciano concluded \cite{4wogu}
that this ``bound appears to rule
out minimal SU(5) with a great desert'' unless the assumptions
underlying the theoretical computation of the proton decay rate were
significantly in error.

By 1987, proton decay bounds had increased, but a more significant
change had taken place.  At the 8th WOGU, Marciano reported \cite{8wogu}
an experimental weak mixing angle determined from a global fit of all
relevant experimental data\footnote{The global fit assumed that
$m_t\!=\! 45$~GeV.  For $m_t\! =\! 175$~GeV, the central
value of $\sin^2\theta_W(m_W)$ 
should be {\it increased} by 0.004 \cite{amaldi}.}
\cite{amaldi}:
\beq \label{e3}
\sin^2\theta_W(m_W) = 0.228\pm 0.0044 \quad [{\rm global~fit}]\,,
\eeq
which is clearly in disagreement with the SU(5) prediction
[eq.~\ref{e1}].  Equivalently, the strong and
electroweak coupling constants do {\it not} unify in the minimal SU(5)
model.  Marciano explained \cite{8wogu} that the new experimental
result [eq.~\ref{e3}] differed from the old result quoted
in eq.~\ref{e2} ``primarily because of more precise deep-inelastic
$\nu_\mu$ scattering data and refinements in the $W^\pm$ and $Z$ mass
determinations.''  The simplest GUT models with the grand desert between
the electroweak and GUT-scales were now decisively ruled out by
experiment.  Of course, one could always add an intermediate scale and
make the GUT model sufficiently complex to reproduce the observed
$\sin^2\theta_W(m_W)$ and to suppress the
proton decay rate below its experimental
bound.  However, in the same paper \cite{8wogu}, Marciano also notes
that the supersymmetric SU(5) model predicts:
\beq
\sin^2\theta_W(m_W) = 0.237^{+0.003}_{-0.004}-{4\alpha\over
15\pi}\ln{\msusy\over m_W},
\eeq
where $\msusy$ characterizes the scale of low-energy supersymmetry
breaking, which ``is in good accord with experiment'' (for $\msusy\sim
1$~TeV).  Of course, with more precise electroweak data obtained at LEP,
SLC, and the Tevatron during the past eight years, the
latter observation has become much sharper (see Section 3.4 for further
discussions).

In 1989 the tenth and last WOGU was held.  The demise of nWOGU was not a
consequence of theorists giving up
on unification.  In fact, during the early 1990s, the
suggestion that unification of couplings was a strong hint for
low-energy supersymmetry received a significant boost from the new
precision data from LEP \cite{susygrand}.  Moreover, low-energy
supersymmetry has the (possibly unique)
potential for providing a theoretical understanding of the
large hierarchy between the electroweak and Planck scales (the latter
only a few orders of magnitude above the GUT-scale), while preserving
the perturbativity of couplings \cite{natural}.  Thus, starting in
1993, a new series
of meetings, the $n$th International Conference on Supersymmetries in
Physics (SUSY-yy)\footnote{Here, yy$=92+n$~(mod~100), where
$n=1,2,\ldots$.}
was established.  Since supersymmetry is believed to be an essential
ingredient in the unification of fundamental interactions and particles,
SUSY-yy is a worthy successor to nWOGU.
Like their predecessors, the organizers of SUSY-yy hold high
expectations for future meetings.  Low-energy supersymmetry implies the
existence of supersymmetric phenomena associated with the electroweak
scale.  New experimental facilities at the Tevatron and LEP-2
have increased their sensitivities,
and could provide the first hints for supersymmetric
particles.  Future experiments at the LHC (and other supercolliders now
under development) have the potential to explore in detail the
properties of supersymmetric particles and their interactions.  These
data could then provide crucial clues to the nature of Planck-scale
physics.

Once again, we have come to Philadelphia to explore the consequences of
unification physics. One hopes that nature will be kinder this second
time around.

\section{DEFINING THE MSSM}

\subsection{The MSSM particle spectrum}

The minimal supersymmetric extension of the Standard Model (MSSM)
consists of taking the Standard Model
and adding the corresponding supersymmetric partners \cite{Haber85}.
In addition, the MSSM contains two hypercharge
\hbox{$Y\!=\!\pm 1$} Higgs
doublets, which is the minimal structure for the Higgs sector of an
anomaly-free supersymmetric extension of the Standard Model.
The supersymmetric structure of the theory also requires (at least) two
Higgs doublets to generate mass for both ``up''-type and ``down''-type
quarks (and charged leptons) \cite{Inoue82,Gunion86}.
All renormalizable supersymmetric interactions
consistent with (global) \hbox{${\rm B}\!-\!{\rm L}$}
conservation (B$=$baryon number and L$=$lepton number) are included.
The latter condition is achieved by employing the following R-parity
conserving superpotential:
\beqa \label{wrparity}
  W & = &\epsilon_{ij} \bigl[ (h_L)_{mn} \widehat H^i_1 \widehat L^j_m
      \widehat E_n + (h_D)_{mn}\widehat H^i_1 \widehat Q^j_m\widehat D_n
       \nonumber \\
     &&\qquad -  (h_U)_{mn} \widehat H^i_2 \widehat Q^j_m\widehat U_n
       - \mu \widehat H^i_1 \widehat H^j_2 \bigr]\,,
\eeqa
where $\epsilon_{ij}=-\epsilon_{ji}$ (with $\epsilon_{12}=1$) contracts
SU(2) doublet fields.  The parameters introduced above are the $3\times
3$ Yukawa
coupling matrices $h_L$, $h_D$ and $h_U$ (with corresponding generation
labels $m$ and $n$) and the Higgs superfield mass parameter, $\mu$.
The gauge multiplets couple to matter multiplets in a manner consistent
with supersymmetry and the SU(3)$\times$SU(2)$\times$U(1) gauge
symmetry. The spectrum of the MSSM is exhibited in Table~1
(generation labels are suppressed).  Note that
Table~1 lists the interaction eigenstates. Particles with the same
SU(3)$\times$U(1)$_{\rm EM}$ quantum numbers can mix.  For example, the
charginos
[$\widetilde\chi^\pm_j$ ($j=1,2$)] are linear combinations of the
charged winos and higgsinos, while the neutralinos [$\widetilde\chi^0_k$
($k=1,\ldots,4$)] are linear combinations of the 
neutral wino, bino and neutral higgsinos.

\renewcommand{\arraystretch}{1.3}
\setlength{\tabcolsep}{0.01in}
\begin{table}[htb]
\centering
\caption{The MSSM Particle Spectrum}
\vskip6pt
\begin{tabular}{ccc}
          &             & Fermionic \\[-5pt]
Superfield& Boson Fields& Partners \\
\hline
\multicolumn{2}{l}{\us{Gauge Multiplets}}& \\[3pt]
$\wh G$&  $g$&     $\wt g$ \\
$\wh V^a$&$W^a$&  $\widetilde W^a$ \\
$\wh V\pri$&     $B$&      $\widetilde B$ \\
\hline
\multicolumn{2}{l}{\us{Matter Multiplets}} & \\[3pt]
$\ds{\wh L\atop \wh E}$&
leptons $\Bigg\{ \ds{\wt L^j\,=\,(\widetilde\nu,\widetilde e^-)_L  \atop
           \ds{\wt E\,=\,\widetilde e^+_R\hphantom{(\nu,_L)}}}$\hfill&
            $\ds{ (\nu,e^-)_L \atop  e^c_L}$
\\[16pt]
$\ds{\wh Q\atop \ds{\wh U \atop \wh D} } $&
  quarks $\left\{\vbox to 27pt{}   \right.
 \ds{ \wt Q^j\,=\,(\widetilde u_L,\widetilde d_L)
  \atop \ds{\wt U\,=\,\widetilde u^*_R\hphantom{,d_L)^f}
  \atop \ds{\wt D\,=\,\widetilde d^*_R\hphantom{,d_L)^f}} } }$\hfill&
    $\ds{(u,d)_L \atop\ds{u^c_L \atop d^c_L}}$
\\ \noalign{\vskip8pt}
$\ds{\wh H_1\atop \wh H_2}$&
     Higgs $\Bigg\{ \ds{ H^i_1 \atop \ds{H^i_2}}$\hbox to 2.0cm{}&
    $\ds{(\wt H^0_1,\wt H^-_1)_L \atop (\wt H^+_2,\wt H^0_2)_L}$
\\ \noalign{\vskip8pt}
\hline
\end{tabular}
\end{table}

The fundamental origin of supersymmetry breaking is unknown.  This
ignorance can be parameterized by adding the most general
soft-supersymmetry-breaking terms \cite{Girardello82} consistent with
gauge invariance and R-parity conservation.  It is here
where most of the new supersymmetric
model parameters reside.  These include three supersymmetry-breaking
Higgs mass parameters, five hermitian $3\times 3$ scalar
squared-mass matrices, three complex $3\times 3$ matrix $A$-parameters
and three (complex) Majorana gaugino masses:  
\beqa \label{softsusy}
 &&\!\!\! V_{\rm soft}  =  m^2_1 |H_1|^2 \!+\! m^2_2|H_2|^2 \!-\! m^2_{12}
                 (\epsilon_{ij} H^i_1H^j_2 + {\rm h.c.}) \nonumber
\\
 &&+ (M^2_{\widetilde Q})_{mn}\,\widetilde Q^{i*}_m
        \widetilde Q^i_n 
   +  (M^2_{\widetilde U})_{mn}\,\widetilde U_m^*\widetilde U_n        
          \nonumber \\
 &&+  (M^2_{\widetilde D})_{mn}\,\widetilde D_m^*\widetilde D_n
            \nonumber \\
 && +  (M^2_{\widetilde L})_{mn}\,\widetilde L^{i*}_m\widetilde L^i_n
      + (M^2_{\widetilde E})_{mn}\,\widetilde E_m^*\widetilde E_n
            \nonumber \\
 && +  \epsilon_{ij} \bigl[ (h_L A_L)_{mn} \widetilde H^i_1
       \widetilde L^j_m \widetilde E_n\! +\! (h_D A_D)_{mn}
       \widetilde H^i_1\widetilde Q^j_m\widetilde D_n \nonumber \\
 &&\quad\qquad - (h_U A_U)_{mn} \widetilde H^i_2
        \widetilde Q^j_m\widetilde U_n + {\rm h.c.}\bigr]\nonumber \\
 && +  \half \left[ M_3\, \widetilde g
   \,\widetilde g + M_2 \widetilde W^a\widetilde W^a
  + M_1 \widetilde B \widetilde B +{\rm h.c.}\right]\,.
\eeqa

\subsection{MSSM-124}

It is instructive to count the number of independent parameters of the
MSSM.  To see how the counting works, consider the 
Standard Model with one complex hypercharge-one Higgs doublet.  The
gauge sector consists of three real gauge couplings ($g_3$, $g_2$ and
$g_1$) and the QCD vacuum angle ($\theta_{QCD}$).  The Higgs sector
consists
of one Higgs squared-mass parameter and one Higgs self-coupling ($m^2$
and $\lambda$).  Traditionally, one trades in the latter two real
parameters for the vacuum expectation value ($v=246$~GeV) and the
physical Higgs mass.  The fermion sector consists of three 
Higgs-Yukawa coupling matrices $h_L$, $h_U$, and $h_D$.  
Initially, $h_L$, $h_U$, and $h_D$ are 
arbitrary complex $3\times 3$ matrices, which in total depend
on 27 real and 27 imaginary degrees of freedom.  

But, most
of these degrees of freedom are unphysical.  In particular,
in the limit where $h_L\!=\! h_U\!=\! h_D\!=\! 0$,
the Standard Model possesses a global U(3)$^5$  symmetry
corresponding to three generations of the five
SU(3)$\times$SU(2)$\times$U(1) multiplets: $(\nu_m$, $e^-_m)_L$,
$(e^c_m)_L$, $(u_m, d_m)_L$, $(u^c_m)_L$, $(d^c_m)_L$, where $m$
is the generation label.  Thus, one can make global U(3)$^5$ rotations
on the fermion fields of the Standard Model to absorb the unphysical degrees
of freedom of $h_L$, $h_U$, and $h_D$.  A U(3) matrix can be
parameterized by three real angles and six phases, so that with the most
general U(3)$^5$ rotation, we can apparently remove 15 real angles
and 30 phases from $h_L$, $h_U$, and $h_D$.
However, the U(3)$^5$ rotations
include four exact U(1) global symmetries of the Standard Model, namely
B and the three separate lepton numbers L$_e$, L$_\mu$ and L$_\tau$.
Thus, one can only remove 26 phases from $h_L$, $h_U$,
and $h_D$.  This leaves 12 real parameters (corresponding to six quark
masses, three lepton masses,\footnote{The neutrinos in the Standard Model
are automatically massless and are not counted as independent degrees of
freedom in the parameter count.} and three CKM mixing angles) and one
imaginary degree of freedom (the phase of the CKM matrix). Adding up
to get the final result, one finds that the Standard Model possesses 19
independent parameters (of which 13 are associated with the flavor
sector).

We now repeat the analysis for the MSSM \cite{savas}.  The gauge sector
consists of four Standard Model real parameters ($g_3$, $g_2$, $g_1$
and $\theta_{QCD}$), and three complex gaugino mass parameters
($M_3$, $M_2$, and $M_1$).  The Higgs sector adds two real squared-mass
parameters ($m_1^2$ and $m_2^2$) and two complex mass parameters
($m_{12}^2$ and $\mu$).  In fact, two of the imaginary degrees of
freedom can be removed.  Consider the limit where
$\mu\!=\! m_{12}^2\!=\! 0$, 
all Majorana gaugino mass parameters are zero and all
matrix $A$-parameters are zero.  The theory in this limit possesses
two flavor-conserving global U(1) symmetries \cite{scott}: a continuous
R symmetry [U(1)$_{\rm R}$] and a Peccei-Quinn symmetry
[U(1)$_{\rm PQ}$].  Thus, one can make
global U(1)$_{\rm R}$ and U(1)$_{\rm PQ}$ rotations on the MSSM fields
to remove two unphysical degrees of freedom from among $\mu$, $m_{12}^2$
and the three complex gaugino Majorana mass parameters (unphysical
degrees of freedom in the matrix $A$ parameters will be addressed
below).  It is convenient to perform a U(1)$_{\rm R}$ rotation
in order to make the gluino mass real and positive ({\it i.e.},
$M_3>0$), followed by a $U(1)_{\rm PQ}$ rotation
to remove a complex phase from $m_{12}^2$.  Since the tree-level Higgs
potential depends on the Higgs mass parameters $m_i^2+|\mu|^2$ ($i=1,2$)
and $m_{12}^2$, it follows that
the tree-level Higgs potential is CP-conserving \cite{Gunion86}.
Thus, three Higgs
sector mass parameters can be traded in (at tree-level) for two real
vacuum expectation values $v_1$ and $v_2$ [or equivalently, $v^2\equiv
v_1^2+v_2^2=(246~{\rm GeV})^2$ and $\tan\beta\equiv v_2/v_1$] and one
Higgs mass [usually taken to be the mass of the CP-odd Higgs scalar
($\ha$)].  The parameters $\tanb$ and $\mha$ can then be used to predict
the masses of the other MSSM Higgs bosons (the CP-even Higgs states
$\hl$ and $\hh$ and a charged Higgs pair $\hpm$) and their
couplings \cite{Gunion86}.
Thus, among the gaugino and Higgs/higgsino mass parameters, there are
seven real
degrees of freedom ($v$, $\mha$, $\tanb$, $M_3$, $|M_2|$, $|M_1|$, and
$|\mu|$) and three phases (${\rm arg}~M_2$, ${\rm arg}~M_1$, and ${\rm
arg}~\mu$).

Finally, we must examine the flavor sector of the MSSM.  In addition
to $h_L$, $h_U$, and $h_D$ of the Standard Model, we have three
arbitrary complex $3\times 3$ matrix $A$-parameters, $A_L$, $A_U$, and
$A_D$ which adds an additional 27 real and 27 imaginary parameters.
Furthermore, the five scalar hermitian $3\times 3$ squared-mass matrices
$M_{\widetilde Q}^2$, $M_{\widetilde U}^2$, $M_{\widetilde D}^2$,
$M_{\widetilde L}^2$, $M_{\widetilde E}^2$ contribute a total of 30
real and 15 imaginary degrees of freedom.  To remove the unphysical
degrees of freedom, we employ global U(3)$^5$ rotations on the {\it
superfields} of the model (thereby preserving the form of the
interactions of the gauginos with matter). This analysis
differs from the Standard Model analysis in that the MSSM possesses
only one global lepton number L. (In particular, L$_e$, L$_\mu$, and
L$_\tau$ are no longer separately conserved in general, for the case
of arbitrary sneutrino masses).  Thus, 
global U(3)$^5$ rotations can remove 15 real parameters
and 28 phases.  Hence, the flavor sector contains 69 real parameters and
41 phases.  Of these, there are nine quark and
lepton masses, three real CKM angles, and 21 squark and slepton masses.
This leaves 36 new real mixing angles to describe the
squark and slepton mass eigenstates and 40 new CP-violating phases that
can appear in squark and slepton interactions!

The final count gives 124 independent parameters for the MSSM of which
110 are associated with the flavor sector.  Of these 124 parameters, 18
correspond to Standard Model parameters, one corresponds to a Higgs
sector parameter (the analogue of the Standard Model
Higgs mass),
and 105 are genuinely new parameters of the model.
Thus, an appropriate name for the minimal supersymmetric extension of
the Standard Model as described above is MSSM-124.

Even in the absence of a fundamental theory of supersymmetry
breaking, one is hard-pressed to regard MSSM-124 as a fundamental
theory.
In particular, the ``minimal'' in MSSM refers to the
minimal particle content and not a minimal parameter count.
Moreover, once 
low-energy supersymmetry is
discovered, one of the main tasks of future experiments will be to
measure as many of the 124 parameters as possible.
Nevertheless, MSSM-124 is
not a phenomenologically viable theory over most of its parameter space.
Among the phenomenologically bad features of this model 
are: (i) no separate
conservation of L$_e$, L$_\mu$, and L$_\tau$; (ii) unsuppressed
flavor changing neutral currents (FCNC's); and (iii) electric dipole
moments of the electron and neutron that are 
inconsistent with the experimental bounds.  As a result,
almost the entire MSSM-124 parameter space is ruled out!
This theory is viable only at very special ``exceptional'' points of
the full parameter space.  MSSM-124 is also theoretically deficient
since
it provides no explanation for the origin of the flavor-sector
parameters
(and in particular, why these parameters conform to the exceptional
points of the parameter space mentioned above).  In addition, no
fundamental
explanation is provided for the origin of electroweak symmetry breaking.

There are two general approaches for treating MSSM-124.  In the
low-energy approach, an attempt is made to elucidate the nature of
the exceptional points in the MSSM-124 parameter space that are
phenomenologically viable.  Consider the following two possible choices.
First, one can arbitrarily assert that
$M_{\widetilde Q}^2$, $M_{\widetilde U}^2$, $M_{\widetilde D}^2$,
$M_{\widetilde L}^2$, $M_{\widetilde E}^2$ and the matrix $A$-parameters
are generation-independent (horizontal universality
\cite{savas,georgi}). Alternatively, one can simply
require that all the aforementioned matrices are flavor diagonal in a
basis where the quark and lepton mass matrices are diagonal (flavor
alignment \cite{seibergnir}).  In either case, L$_e$, L$_\mu$ and
L$_\tau$ are separately
conserved, while tree-level FCNC's are automatically absent.  Of course,
the number of free parameters characterizing the MSSM in either of
these two cases is substantially less than 124.  Both
scenarios are phenomenologically viable.  However, such approaches are
almost certainly too restrictive.  First, the phenomenologically viable
region of MSSM-124 parameter space is surely larger than that of these
two scenarios.  Second, it is likely that there is no fundamental
theory of supersymmetry breaking that precisely produces either scenario
above.  Nevertheless, one could reasonably hope that one these two models
might serve as useful
first approximations to the correct theory.  Of course, the
deviations of the correct theory from the above approximations
would contain critical clues for the origin of the flavor structure
of the MSSM.

In the high-energy approach, one
treats the parameters of the MSSM as running parameters and imposes a
particular structure on the soft supersymmetry breaking terms at
a common high energy scale [such as the Planck
scale ($\mpl$) or GUT scale ($\mx$)].
Using the renormalization group equations (RGEs), one can then
derive the low-energy MSSM parameters.  This approach is usually
characterized by the mechanism in which supersymmetry
breaking is communicated to the effective low energy theory.   Two
theoretical scenarios have been examined in detail:
gravity-mediated and gauge-mediated supersymmetry breaking.
One bonus of such approaches is that one of the diagonal Higgs squared-mass 
parameters is typically driven negative by renormalization group
evolution.  Thus, electroweak symmetry breaking is generated
radiatively, and the resulting electroweak symmetry breaking scale is
intimately tied to the scale of low-energy supersymmetry breaking.

A truly Minimal SSM does not (yet) exist.  The MSSM
particle content must be supplemented by assumptions about the origin of
supersymmetry-breaking that lie outside the low-energy domain of the
model.  Moreover, a comprehensive map of the
phenomenologically acceptable region of MSSM-124 parameter space does
not yet exist. This
presents a formidable challenge to supersymmetric particle searches that
must impose some parameter constraints while trying to ensure that the
search is as complete as possible.

\subsection{The minimal-SUGRA-inspired MSSM}

Consider a supergravity (SUGRA) theory consisting of two sectors: a
``hidden'' sector,\footnote{A hidden sector
consists of fields that carry no SU(3)$\times$SU(2)$\times$U(1)
quantum numbers and do not have any renormalizable interactions
with the MSSM fields.}
in which supersymmetry is spontaneously broken and a
``visible'' sector consisting of the MSSM fields.  Because of the
gravitational interactions that necessarily couple the two
sectors, the effects of the hidden sector supersymmetry breaking will
be transmitted to the MSSM.  One finds that the resulting low-energy
effective theory below the Planck scale consists of the unbroken MSSM
plus all possible soft supersymmetry breaking terms \cite{Weldon}.
In a {\it minimal} SUGRA framework \cite{Nilles}, the soft-supersymmetry
breaking parameters at the Planck scale take a particularly
simple form in which the scalar squared masses
and the $A$-parameters are flavor diagonal and universal \cite{Lykken}:
\beqa \label{plancksqmasses}
 &&M^2_{\wt Q} (\mpl) = M^2_{\wt U}(\mpl) = M^2_{\wt D}(\mpl) =  m_0^2
            {\bf 1}
\,,\nonumber\\
 &&M^2_{\wt L}(\mpl) = M^2_{\wt E}(\mpl) = m_0^2 {\bf 1} \,,\nonumber \\
 &&m^2_1(\mpl) = m^2_2(\mpl) = m_0^2 \,,\nonumber \\
 && A_U(\mpl) = A_D(\mpl) = A_L(\mpl) = A_0 {\bf 1}\,.
\eeqa

In addition, the gauge couplings and gaugino mass parameters are
assumed to unify at the grand unification scale, $\mx$.
Note that this implies that:
\beqa  \label{gunif}
c_1 g_1(M_X) = g_2(M_X) &=& g_3(M_X) = g_U\,, \nonumber \\
M_1(M_X) = M_2(M_X) &=& M_3(M_X) = m_{1/2}\,.
\eeqa
where $c_1\equiv\sqrt{5/3}$ ensures proper normalization of the
U(1)$_{\rm Y}$ coupling constant.  Eq.~(\ref{gunif}) implies that the
low-energy gaugino mass parameters satisfy:
\beq \label{gauginomassrelation}
 M_3 = {g^2_3\over g_2^2} M_2,\qquad M_1 = {5\over 3}\tan^2\theta_W
M_2\,.
\eeq

Finally, $\mu$, and $m_{12}^2$ and the gaugino mass parameters
are assumed (rather arbitrarily) to be initially real at the high scale.
The minimal SUGRA-inspired MSSM has been sometimes called the
constrained MSSM (or CMSSM \cite{cmssm}); I will adopt this nomenclature
in this paper.  It is easy to count
the number of free parameters of the CMSSM.  The low-energy values of
the MSSM parameters are determined by the MSSM RGEs and
the above initial conditions [eqs.~(\ref{plancksqmasses}) and
(\ref{gunif})].  Gauge coupling unification yields a prediction for one
of the gauge couplings in terms of the other two.  In this case, it is
convenient to take $g_1$ and $g_2$ as input, and $\alpha_s\equiv
g_3^2/(4\pi)$ as a prediction.  To the extent that gauge coupling
unification is successful, the number of parameters of the Standard
Model is reduced by one (since $\mx$ is also {\it a priori} a free
parameter).  To be conservative, it is useful to introduce an additional
parameter that reflects possible non-trivial thresholds at the GUT
scale, which could lead to slight changes in the prediction for
$\alpha_s(\mz)$ \cite{bmp}.  In this case,
the Standard Model parameter count would remain unchanged.

Thus, the number of
parameters of the CMSSM are: 18 Standard Model parameters (excluding the
Higgs mass), $m_0$, $m_{1/2}$, $A_0$, $\tanb$, and ${\rm sgn}(\mu)$ for
a total of 23 parameters.  Note that $m_{12}^2$ and $\mu^2$ were traded
in for $v^2$ (which is counted as one of the 18 Standard Model
parameters) and $\tanb$. In this procedure the sign of
$\mu$ is not fixed, and so it remains an independent degree of freedom.
It is tempting to rename this theory MSSM-23.

Clearly, MSSM-23 is much more predictive than MSSM-124.
In particular, one has only four genuinely
new parameters beyond the Standard Model
(plus a two-fold ambiguity corresponding to the sign of
$\mu$), from which one can predict the entire MSSM spectrum and its
interactions.\footnote{In some regions of CMSSM parameter space, infrared
fixed point behavior reduces the number of new parameters even
further \cite{fixedpoint}.}
The disadvantage of the CMSSM is that
the theoretical motivation underlying the initial conditions given in
eq.~(\ref{plancksqmasses}) is rather weak.  Although these initial
conditions correspond to a minimal SUGRA framework (specifically,
the kinetic energy terms for the gauge and matter fields are assumed
to take a minimal canonical form), there is no theoretical principle
that enforces such a minimal structure.  In fact, it is now generally
believed that supergravity-based (or superstring-based) supersymmetry
breaking theories generically predict non-universal scalar
masses \cite{nonuniv}.

A number of attempts have been made to perturb the CMSSM initial
conditions \cite{cmssmpert,snowtheory2}
in a phenomenologically viable manner ({\it e.g}, without
generating dangerous FCNC's \cite{choudhury}).  For example, one can
introduce separate mass scales
for the Higgs and squark/slepton soft-supersymmetry-breaking masses.
One can also introduce non-universal scalar masses, but
restrict the size
of the non-universal terms to be consistent with phenomenology.
String-inspired models have
provided an interesting parameterization of the deviation from
universality of the soft-supersymmetry-breaking parameters
\cite{nonuniv}. Both
``bottom-up'' and ``top-down'' approaches have been useful in studying
the possible form for supersymmetry breaking at the low-energy and
high-energy scales.

Finally, although gaugino mass unification [see
eq.~(\ref{gunif})] is an integral part of the CMSSM initial
conditions, it is {\it not} required by phenomenological constraints.
Thus, it is also of interest to consider the phenomenological
consequences of non-universal gaugino masses.  For example,
the phenomenology of $M_1\simeq M_2$ [in contrast
to $M_1\simeq 0.5 M_2$ predicted by eq.~(\ref{gauginomassrelation})] has
recently been advocated \cite{kane} in order to provide a possible
explanation for the famous CDF $ee\gamma\gamma$ event.

\subsection{Models of gauge-mediated supersymmetry breaking}

In an alternative to the SUGRA approach, the theory of
gauge-mediated supersymmetry breaking posits that
supersymmetry breaking is transmitted to the MSSM via gauge forces.
The canonical structure of such models involves a hidden sector
where supersymmetry is broken, a ``messenger'' sector consisting of
messenger fields with SU(3)$\times$SU(2)$\times$U(1) quantum
numbers, and a sector containing
the fields of the MSSM \cite{dine,kolda}.  The
direct coupling of the messengers to the hidden sector generates a
supersymmetry-breaking spectrum in the messenger sector.
Finally, supersymmetry
breaking is transmitted to the MSSM via the virtual exchange of the
messengers.

In models of gauge-mediated supersymmetry breaking,
scalar squared-masses are expected to be flavor independent since
gauge forces are flavor-blind.  In the simplest models, there is
one effective mass scale, $\Lambda$, that determines all low-energy
scalar and gaugino mass parameters through loop-effects
(while no $A$-parameters are generated).
In order that the resulting superpartner masses
be of order 1~TeV or less, one must have $\Lambda\sim 100$~TeV.
The generation of $\mu$ and $m_{12}^2$ lies outside the ansatz of
gauge-mediated supersymmetry breaking.  The initial conditions for the
soft-supersymmetry-breaking running parameters are fixed at the
messenger scale $M$, which characterizes the average mass of
messenger particles.  In principle, $M$ can lie anywhere between
(roughly) $\Lambda$ and $10^{16}$~GeV (in models with larger values of
$M$, supergravity-mediated effects would dominate the gauge-mediated
effects).
Thus, the minimal gauge mediated model (MGM) \cite{thomas} contains 18
Standard Model parameters,
$\Lambda$ (which determines the supersymmetric scalar and gaugino
masses), and $\tanb$ and ${\rm arg}(\mu)$ [after trading in 
$m_{12}^2$ and $|\mu|^2$
for $v$ and $\tanb$].  There is also a weak logarithmic
dependence on $M$, which enters through RGE running.
We thus end up with 22 free parameters, which implies that
the MGM is even more predictive than the CMSSM.
However, the MGM is not a fully realized model.  The sector of
supersymmetry-breaking dynamics can be very complex, and it is fair to
say that no simple compelling model of gauge-mediated supersymmetry yet
exists.  Nevertheless, this is an area of intense theoretical activity,
and it will be interesting to see the variety of MSSM's that emerge from
this approach over the next few years.

\section{PHENOMENOLOGICAL ISSUES}

\subsection{Can low-energy supersymmetry be excluded?}

Supersymmetric particles have not yet been discovered.\footnote{A
few intriguing experimental anomalies have encouraged various
supersymmetric interpretations \cite{kane,hera}.  In most cases,
data now being collected at LEP-2 will
either provide substance to such claims or rule them out.
Data from Run-II of the Tevatron (which is scheduled to begin in 1999)
can also provide the important corroborating evidence.}
Thus, direct searches for supersymmetric particles at colliders have so
far provided lower bounds for supersymmetric particle masses.
These results are summarized by the Particle Data Group \cite{pdgbook}.

Can the MSSM be ruled out if no supersymmetric particles are discovered
at future colliders?  It is generally believed that low-energy
supersymmetric theories require that supersymmetric particle masses
should be less than ${\cal O}(1~{\rm TeV})$.  The argument follows from
the naturalness requirement that is invoked to explain the existence of
the large hierarchy between the electroweak scale and the Planck scale.
This hierarchy is unnatural in the Standard Model, since there is no
natural mechanism for keeping scalar masses light.  Supersymmetry can
naturally incorporate light scalars by relating them to light fermions
which can have
small masses due to weakly broken chiral symmetries.  

If the scale of
supersymmetry breaking is of order 1~TeV or less, then a Higgs mass of
order the electroweak scale is still natural.  However, as the
supersymmetry-breaking scale increases, the condition for the
fine-tuning of
parameters (in order to keep the Higgs mass light) becomes more severe.
Theorists have attempted to quantify the ``degree of naturalness'', and
thereby deduce upper bounds for supersymmetric particle masses (see,
{\it e.g.}, Refs.~\cite{barbieri,anderson}).  The naturalness
conditions obtained are somewhat arbitrary, as are the corresponding
conclusions.  Although some interesting observations can be made,
it is not possible to obtain rigorous upper limits on
supersymmetric particle masses.  
Personally speaking, I am willing to concede that if supersymmetric
particles are not discovered
at the LHC, then supersymmetry is not relevant for explaining the origin
of the electroweak scale.

\subsection{The MSSM Higgs Sector}

There is one case in the MSSM where a particle mass upper limit can be
rigorously obtained.
The mass of the light CP-even neutral Higgs boson, $h^0$, in the MSSM
can be calculated to arbitrary accuracy in terms of two parameters of
the Higgs sector, $m_{A^0}$ and $\tan\beta$, and other MSSM
soft-supersymmetry-breaking parameters that affect the Higgs mass
through virtual loops~\cite{hhprl}.  If the
scale of supersymmetry breaking is much larger than $m_Z$, then large
logarithmic terms arise in the perturbation expansion.  These large
logarithms can be resummed using renormalization group (RG) methods.
The logarithmic sensitivity to the supersymmetry breaking scale implies
that the Higgs mass upper bound depends only weakly on the choice
of an upper bound for supersymmetric particle masses.

At tree level, the MSSM predicts that $\mhl\leq\mz|\cos 2\beta|\leq\mz$.
If this prediction were accurate, it would imply that the Higgs boson
must be discovered at the LEP-2 collider (running at a
center-of-mass energy of 192~GeV, with an integrated luminosity of
300 ${\rm pb}^{-1}$).  Absence of a Higgs boson lighter than $\mz$ would
naively rule out the MSSM.  When radiative corrections are included, the
light Higgs mass upper bound is increased significantly.  In the
one-loop leading logarithmic approximation \cite{hhprl},
\begin{equation} \label{mhlapprox}
\mhl^2\lsim\mzz\cos^2 2\beta+{3g^2 m_t^4\over
8\pi^2\mww}\,\ln\left({M^2_{\tilde t}\over \mt^2}\right)\,,
\end{equation}
where $M_{\tilde t}$ is the (approximately) common mass of the
top-squarks.  Observe that the Higgs mass upper bound is very sensitive
to the top quark mass and is logarithmically sensitive to the
top-squark masses.  Although eq.~(\ref{mhlapprox}) provides a rough
guide to the Higgs mass upper bound, it is certainly insufficient for
Higgs searches at LEP-2, where the Higgs mass reach depends delicately on the
MSSM parameters.  Moreover, in order to compare precision Higgs
measurements with theory, a more precise expression
for the Higgs mass is needed. The formula for the full one-loop
radiative corrected Higgs mass has been obtained in the literature,
although it
is very complicated since it depends in detail on the virtual
contributions of the MSSM spectrum~\cite{honeloop}.
If the supersymmetry breaking scale is larger than a few
hundred GeV, then RG methods are essential for summing up the effects of
the leading logarithms and obtaining an accurate prediction.

The computation of the RG-improved
one-loop corrections requires numerical integration of a coupled set of
RGEs~\cite{llog}. (The dominant two-loop next-to-leading
logarithmic results are also known~\cite{hempfhoang}.)
Although this program has been
carried out in the literature, the procedure is unwieldy
and not easily amenable to large-scale Monte-Carlo analyses.
Recently, two groups have presented simple analytic procedures for
accurately approximating $m_{h^0}$.
These methods can be easily implemented, and incorporate both the
leading one-loop and two-loop effects and the RG-improvement.
Also included are the leading effects at one loop of supersymmetric
thresholds (the most important effects of this type are squark mixing
effects in the third generation). Details of the techniques can
be found in Refs.~\cite{hhh} and \cite{carena}, along with other
references to the original
literature.  Here, I shall quote two specific bounds, assuming
$\mt\!=\! 175$~GeV and $M_{\tilde t}\!\lsim\! 1$~TeV:
$\mhl\lsim 112$~GeV if top-squark mixing is negligible, while
$\mhl\lsim 125$~GeV if top-squark mixing is ``maximal''.
Maximal mixing corresponds to
an off-diagonal squark squared-mass that produces the largest value of
$\mhl$.  This mixing leads to an extremely large splitting of top-squark
mass eigenstates.  A more realistic choice of top-squark parameters
leads to a Higgs mass upper bound of about 120~GeV.

\subsection{Implications of precision electroweak data}

Virtual supersymmetric particle exchange can influence many
experimental processes.
After eight years of precision electroweak data from LEP, SLC, and the
Tevatron, the Standard Model predictions for many observables have been
confirmed with remarkable accuracy.  The LEP Electroweak Working Group
(LEPEWWG) continues to update its fits of precision electroweak data.
In its most recent work \cite{ewwg},
a Standard Model global fit to 21 electroweak observables is
presented.  Only two observables exhibit a pull of two standard
deviations or greater, while the $\chi^2$/d.o.f. for the fit is 17/15.
Thus, the precision electroweak data shows no significant deviation from
Standard Model expectations.  There is some sensitivity to the Standard
Model Higgs mass via its virtual effects.  The result obtained from the
global fit is $\mhl=115^{+116}_{-66}$~GeV, or $\mhl<420$~GeV at 95$\%$
CL.  These results are relevant for the MSSM in the following sense.
The MSSM is a decoupling theory.  If all supersymmetric particle masses
are large compared to $\mz$, then the virtual effects of supersymmetric
particle exchange decouple from electroweak observables measured at an
energy scale of order $\mz$ or below.  Moreover, if $\mha\gg\mz$, then
the effects of the non-minimal Higgs bosons $\hh$, $\ha$ and $\hpm$ also
decouple, while $\hl$ remains light ($\mhl\lsim 125$~GeV).
In the decoupling limit, the light CP-even Higgs
boson $\hl$ has Standard Model coupling strengths to the Standard Model
fermions and gauge bosons, and in this sense is indistinguishable from
the Higgs boson of the Standard Model \cite{declimit}.  Thus, as long as
all supersymmetric particles (and the non-minimal Higgs bosons) are
sufficiently heavy (in practice, masses above 200~GeV are sufficiently
decoupled), then the MSSM\footnote{We still must make the
additional assumption
that the appropriate region of MSSM-124 parameter space has been
selected to avoid the flavor problems discussed in
Section 2.2.  In principle, the flavor problem
could be solved by the decoupling properties of the
MSSM.  However, the strongest constraints that exist for
FCNC processes would require first and second generation squark and slepton
masses to be above about 50~TeV
in the absence of any other FCNC suppression mechanism \cite{dkl}.}
provides an equally good description of
the precision electroweak data {\it as long as the Higgs
boson mass obtained in the Standard Model global fit is consistent with
the MSSM light
Higgs mass upper bound}.  This latter condition is indeed satisfied by
the LEPEWWG global fit.

If some supersymmetric
particles are light (say, below 200~GeV but above present experimental
bounds), then it is possible that the MSSM could either improve or
destroy the LEPEWWG global fit.  A few years ago, when the rate for
$Z\to b\bar b$ was four standard deviations above the Standard Model
prediction, the possibility that the MSSM could improve the LEPEWWG fit
was taken quite seriously.  However, it is
hard to imagine that the MSSM could substantively improve the 
present LEPEWWG fit (given
the goodness of the Standard Model fit in comparison to an MSSM fit,
which necessarily involves more degrees of freedom). On the other hand,
the MSSM could significantly decrease the goodness of the fit.  This
possibility has been explored recently in Ref.~\cite{erler}.  It was
shown that there exists a range of parameters of the CMSSM and the
MGM in which all supersymmetric particle masses are above their direct
search bounds, but the global fit of electroweak data is significantly
worse than the corresponding Standard Model fit.  Thus, the allowable
CMSSM and MGM parameter spaces are slightly smaller than the regions
ruled out by the direct supersymmetric particle searches.

\subsection{Supersymmetric unification revisited}

Electroweak observables are also sensitive to the strong coupling
constant through the QCD radiative corrections.  The LEPEWWG global fit
extracts a value of $\alpha_s(\mz)=0.120\pm 0.003$, which is in good
agreement with the world average of $\alpha_s(\mz)=0.118\pm 0.003$
quoted by the Particle Data Group \cite{pdgbook}.  Thus, previous claims
that $\alpha(\mz)\simeq 0.11$ \cite{shifman} now seem somewhat
disfavored.  This result has important implications for the viability of
supersymmetric unification.  In Section 1, I briefly reviewed the
history of unification models by focusing on the prediction of
$\sin^2\theta_W$.  Given the $\sin^2\theta_W$ is so well measured at LEP
and SLC, it makes more sense to use this as input data (along with the
fine-structure constant).  This data can be used to obtain accurate
values for $g_1(\mz)$ and $g_2(\mz)$ [in either the $\overline{\rm MS}$
or $\overline{\rm DR}$ schemes].  Extrapolating to high energies using
either Standard Model or MSSM RGEs (with appropriate treatment of the
low-energy supersymmetric thresholds), one then defines the mass scale
at which $c_1 g_1$ [$c_1=\sqrt{5/3}$] and $g_2$ meet as the unification
scale, $\mx$.  If the strong coupling constant $g_3$ also
coincides with $c_1 g_1$ and $g_2$ at $\mx$,
then by extrapolating back down to $m_Z$, one obtains a
prediction for $\alpha_s(\mz)$.  The result of this
exercise for the CMSSM is 
$\alpha_s(\mz)>0.126$ \cite{bmp},\footnote{The corresponding result for the
Standard Model extrapolation, $\alpha_s(\mz)\simeq 0.073\pm 0.002$
\cite{Langacker}, is of course
many standard deviations away from the experimentally observed result.}
assuming that all squark masses are below 1~TeV (similar results have
been obtained in Ref.~\cite{polonsky}).

Thus, naive unification of gauge couplings in the MSSM does not quite
work.
However,  the results of the analysis quoted above do not include the
effects of possible high
energy thresholds generated from the spectrum of masses of superheavy
GUT particles.  Taking such effects into account could either improve
the situation or make it worse, depending on the details of the model.
Two examples taken from the recent literature exhibit some of the
possibilities.  Consider GUT models employing the missing
doublet mechanism to solve the doublet--triplet splitting problem ({\it
i.e.}, why are the Higgs weak doublets so much lighter than the
superheavy Higgs triplets that typically occur in GUT models).
It has been shown \cite{bmp,dedes} that there
is a range of the model parameter space  
where the heavy threshold
corrections are sufficient to lower $\alpha_s(\mz)$ to a
value in agreement with experimental observation.
In  Ref.~\cite{raby}, some
SO(10) grand unified models are studied which also exhibit
an acceptable prediction for $\alpha_s(\mz)$ as a result of high energy
threshold corrections.   In addition, interesting correlations are found
between the predicted value of $\alpha_s(\mz)$ and the proton lifetime.

\subsection{Search for supersymmetry at future colliders}

The search for supersymmetry at future colliders presents some important
challenges.  Perhaps the first order of business is to discover the
light CP-even Higgs boson.  If no Higgs boson with mass
below about 125~GeV is
discovered, then MSSM-124 is ruled out.  LEP-2 will eventually be
sensitive to Higgs masses up to about 100~GeV.  To close the gap
completely, one must first look to the hadron colliders.  It has recently been
pointed out that an upgraded Tevatron (with luminosity a factor of ten
larger than the Main Injector) has the potential to detect Higgs bosons
with masses up to about 130~GeV \cite{gunion}.  Whether all the machine
and detector requirements can be met to reach this goal remains to be
seen.
For the LHC, the designs of the ATLAS and CMS detectors are being
optimized for a discovery of a Higgs boson
with a mass between 90 and 130 GeV via its
rare $\gamma\gamma$ decay mode (the expected branching ratio is about
$10^{-3}$ in the Standard Model) \cite{lhchiggs}.
To achieve success in the Higgs
search via the $\gamma\gamma$ mode at the LHC will require high
electromagnetic
calorimeter resolution (at about the 1\% level) and maximal luminosity.
At present, the LHC coverage of the MSSM
Higgs sector parameter space is nearly complete, although
small gaps in the MSSM parameter space may still
exist.  Because the search techniques in some cases depend on the
observation of small signals above
significant Standard Model backgrounds, it may be difficult to
definitively rule out the MSSM if
no Higgs signal is observed at the LHC.

If a very high-energy (next) linear $e^+e^-$
collider (NLC) is built, then it will be able to extend the LEP-2 Higgs
search up to a Standard Model Higgs mass of about 
350 (800)~GeV for $\sqrt{s}\!=\! 500~(1000)$~GeV \cite{hehsnow}.  With
$\sqrt{s}\gsim
300$~GeV and a total integrated luminosity of $\gsim 1$~fb$^{-1}$, the
NLC would either discover $\hl$ or rule out the MSSM.  Note that
although the discovery of $\hl$ is essential for the MSSM to remain
viable, the discovery does not serve to confirm the MSSM.  For example,
if the other non-minimal Higgs bosons are significantly heavier than
the $Z$, then the properties of the $\hl$ will be indistinguishable from
the Standard Model Higgs boson.

Thus, the ultimate confirmation of low-energy supersymmetry requires the
direct discovery of supersymmetric particles.  A
comprehensive experimental program
to study low-energy supersymmetric phenomena must accomplish four goals:
(i) initial discovery of the supersymmetric particles; (ii)
verification of the supersymmetric structure of their interactions;
(iii) identification of the low-energy supersymmetric spectrum and its
symmetries [{\it e.g.}, MSSM or beyond, R-parity conservation or
violation]; and
(iv) measurement of the low-energy supersymmetry model parameters.
For example, in the case of the MSSM, step (iv) would consist of
measuring as many of the MSSM-124 parameters as possible.  One would
then make use of these experimental measurements to determine whether
these parameters approximately matched the expectations of particular
models such as the CMSSM or the MGM.

This program is highly non-trivial.  In developing strategies
for supersymmetric particle searches, one is tempted to look for
shortcuts.  For example, devising a strategy for discovery and
exploration of the CMSSM (a.k.a.~MSSM-23) is clearly a simpler task than
the corresponding strategy for MSSM-124.  Even well established
phenomenology, such as the missing-energy signal generated by the
escaping
lightest supersymmetric particle (LSP), can be altered by a change of
assumptions.\footnote{For example, consider the model recently
proposed in
Ref.~\cite{raby2}, in which the gluino is the LSP.}  Another example of
a well known fact of canonical supersymmetric phenomenology---most
supersymmetric decays do not involve photons---does not 
necessarily apply to
models of gauge-mediated supersymmetry breaking
\cite{thomas,others}.\footnote{Ref.~\cite{kane}
reminds us that photons can be also
produced with large branching ratio in neutralino decay,
$\widetilde\chi^0_2\to\widetilde\chi^0_1\gamma$
\cite{wyler}, in models where the electroweak gaugino mass parameters
are approximately equal ($M_2\simeq M_1$).  This is neither the CMSSM
nor the
MGM, but another interesting point in MSSM-124 parameter space.}
In these models the gravitino ($\widetilde g_{3/2}$)
is the LSP.   The decays of a supersymmetric particle will eventually
produce the next-to-lightest supersymmetric particle
(NLSP), which is typically the lightest neutralino
($\widetilde\chi^0_1$). Over a significant fraction of the model
parameter space, $\widetilde\chi^0_1$ decays inside the detector to
$\gamma+\widetilde g_{3/2}$. In this example,
{\it all} supersymmetric particle decay chains would contain a photon.

Perhaps a non-minimal model of
low-energy supersymmetry is the correct theory of TeV-scale
physics.  The phenomenology of such models could be
considerably different from the canonical supersymmetric phenomenology
usually analyzed.  For example, in R-parity violating models, the LSP
can decay into visible matter.  Here is an example where the missing
energy signal could be irrelevant for low-energy supersymmetry searches.
The challenge for supersymmetry searches at
future colliders is to allow for all possible phenomenological scenarios.

\section{BEYOND MSSM-124}

In this paper, I have attempted to restrict the definition of the MSSM
to the properties of the effective low-energy theory below the 1~TeV
energy scale.
I argued that the most general approach then leads to MSSM-124 which
possesses a viable phenomenology only at exceptional points in its
parameter space.  More precisely, the phenomenology is viable at these
exceptional points, and in the regions of parameter space defined by
small perturbations around these exceptional points.  Presumably these
perturbations are generated by new physics at higher energy
scales.  Such perturbations can generate rare processes that lie
outside the Standard Model (as well as outside the typical MSSM).
Possible consequences include: proton decay, L$_e$, L$_\mu$ and/or
L$_\tau$ violation, L violation (leading, {\it e.g.},
to neutrino masses), and new sources of CP-violation.  For
example, in Ref.~\cite{barhall} it was argued that supersymmetric grand
unification models should typically
generate $\mu\to e\gamma$, at a level that may be observable at future
high precision experiments.

\subsection{Supersymmetric models with non-zero neutrino masses}

There
is some evidence for very small but non-zero neutrino masses.  Thus, it is
of interest to
consider a supersymmetric generalization of an extended Standard
Model that contains nonzero neutrino masses.
In the supersymmetric extension of the see-saw model of
neutrino masses \cite{seesaw,grosshab},
one introduces a right-handed neutrino superfield $\widehat
N$, with the following new terms in the superpotential:
\beq  \label{wnu}
\delta W=-\epsilon_{ij}h_N \widehat H^i_2 \widehat L^j\widehat N
         -\half M\widehat N\widehat N\,,
\eeq
where generation labels have been suppressed.
Here, $M$ is the scale of the right-handed neutrino.
In addition, one adds new soft supersymmetry breaking terms:
\beqa \label{softsnu}
\delta  V_{\rm soft}  & = & m^2_N \widetilde N^*\widetilde N +
                 (m_{NN} \widetilde N\widetilde N + {\rm h.c.})
\nonumber \\
 &&- \epsilon_{ij} \left[ h_N A_N \widetilde H^i_2 \widetilde
      L^j \widetilde N+ {\rm h.c.}\right]\,.
\eeqa
Note that the supersymmetric see-saw model
conserves R-parity since lepton number is violated by two units.

For simplicity, consider the one-generation case.  Let $m_D\equiv h_N
v_2/\sqrt{2}$.  If $m_D\ll M$, then the fermion spectrum contains
a very heavy neutrino with mass of order $M$ and a very light neutrino
with mass of order $m_D^2/M$.  This is the see-saw; an appropriately
large choice for $M$ can naturally lead to neutrino masses in the eV
range and below.  In the supersymmetric model, the $\Delta L=2$
interaction responsible for neutrino mass will also generate
sneutrino-antisneutrino mixing.  If CP is conserved, the
$\snu$--$\overline{\snu}$
mixing angle is $45^\circ$, producing a CP-even and
CP-odd scalar mass eigenstate.  
The mass splitting of these two states is of order
the light neutrino mass (although enhancements of $10^3$ are possible
as shown in Ref.~\cite{grosshab}).  In favorable model circumstances,
this small mass difference could be measured in sneutrino pair
production at $e^+e^-$ colliders by detecting
like-sign di-leptons from sneutrino decays (in analogy with the
$B^0$--$\overline {B^0}$ system).

\subsection{New gauge and matter multiplets at the TeV-scale}

Models of neutrino masses necessarily add new structure beyond the
Standard Model.  In the see-saw example, this new structure lives at a
high energy scale.  However, when the physics associated with the high
scale is integrated
out, there is a non-trivial remnant in the effective low-energy theory.
A second possible approach is to add new structure beyond the Standard
Model at the 1~TeV scale.  The supersymmetric extension of such a theory
would be a non-minimal extension of the MSSM.  Possible new structures
include \cite{rizzo}: (i) an enlarged electroweak gauge group beyond
SU(2)$\times$U(1); (ii) the addition of new, possibly exotic, matter
multiplets [{\it e.g.}, a vector-like color triplet with electric charge
(1/3)$e$; such states sometimes occur as low-energy remnants in E$_6$
GUT
models]; or (iii) the addition of low-energy SU(3)$\times$SU(2)$\times$U(1)
singlets.  A possible theoretical motivation for such new structure
arises from the study of phenomenologically viable string
theory ground states \cite{dienes}.

\subsection{The next-to-minimal supersymmetric model}

The next-to-minimal supersymmetric
extension of the Standard Model (NMSSM) consists of adding one
complex singlet Higgs superfield to the MSSM \cite{nmssm}.
This example provides an instructive
case study of a non-minimal supersymmetric Higgs sector.  In
particular, it was noted in Section 3.2 that the experimental absence of
a light CP-even Higgs boson with $\mhl\lsim 125$~GeV would rule out the
MSSM.  In the NMSSM,
the Higgs mass bound depends on an extra assumption beyond
the physics of the low-energy effective theory.  Specifically,
the addition of the Higgs singlet adds a new Higgs self-coupling
parameter $\lambda$ to the theory.\footnote{In contrast,
all Higgs self-couplings of the MSSM are related by supersymmetry to
gauge couplings.  This is the origin of the MSSM bound: 
$\mhl\lsim{\cal
O}(\mz)$.}  The mass of the lightest
neutral Higgs boson can be
raised arbitrarily by increasing the value of $\lambda$ (analogous to
the behavior of the Higgs mass in the Standard Model).  However, if
$\lambda$ is taken to be too large, then perturbation theory
becomes unreliable.  If one imposes the condition that
all couplings remain perturbative up to the Planck scale,  one
finds that at least one Higgs boson of the model must be
lighter than about 150~GeV.

Could the Higgs bosons of the NMSSM escape detection at future
colliders?  Even though a light Higgs state must exist (under the
perturbativity assumption introduced above),
it may be very weakly coupled to quarks, leptons and gauge
bosons if it is primarily composed of the singlet component.  Thus, a
detailed analysis is required to see whether the Higgs search at the NLC
is sensitive to all regions of the NMSSM Higgs sector parameter space.
The analysis of Ref.~\cite{KOT} demonstrates that even for
$\sqrt{s}=300$~GeV, the NLC search would easily detect at least one
Higgs state of the NMSSM.

A similar question can be posed in the case of the LHC Higgs search.
As mentioned in Section 3.5,
the LHC search has nearly complete coverage of the MSSM Higgs
sector parameter space.  It is likely that further development of search
techniques (and improvements of detector technology such as efficient
$b$-tagging) will be able to
close any final loopholes.  Nevertheless, the LHC search is
operating ``at the edge'' of its capabilities.  By relaxing some of the
MSSM constraints to Higgs sector parameters, we expect some holes to
develop in the region of supersymmetric
parameter space accessible to the LHC Higgs search.
Ref.~\cite{GHM} examined this question in detail for the case of
the NMSSM, and concluded that although the region of
inaccessibility is
not large, it is possible to find regions of NMSSM Higgs parameter space
in which no Higgs boson state could be discovered at the LHC.  This
analysis does suggest the possibility that future improvements in search
strategies and detector capabilities may be able to close these
loopholes as well.

\subsection{R-parity violating supersymmetry}

So far, all non-minimal supersymmetric models considered here
have retained R-parity as
a discrete symmetry.  R-parity is a {\bf Z}$_2$ symmetry that
distinguishes between Higgs and matter superfields.  For a particle of
spin $S$, the R-parity quantum number is $R=(-1)^{3(B-L)+2S}$, so
R-parity implies the conservation of \hbox{${\rm B}\!-\!{\rm L}$}.
Consider the MSSM,
but now relax the constraint of R-parity conservation \cite{dreiner}.
In the absence of R-parity
conservation, new terms can be added to the superpotential:
\beqa \label{nrsuppot}
 W_{\rm NR} &=& \epsilon_{ij} \bigl[ (\lambda_L)_{pmn} \widehat L^i_p
      \widehat L^j_m
      \widehat E_n + (\lambda_L^\prime)_{pmn}\widehat L^i_p \widehat
          Q^j_m\widehat D_n \nonumber \\
        && \quad-  \mu_p^\prime \widehat L^i_p \widehat H^j_2 \bigr]
          +(\lambda_B)_{pmn}\widehat U_p \widehat D_m \widehat D_n\,.
\eeqa
New R-parity violating soft supersymmetry breaking terms can also be
obtained by the usual procedure of replacing the superfields of the
superpotential with their
corresponding scalar partners, and introducing a new matrix
coefficient for each term.  Note that the term in eq.~(\ref{nrsuppot})
proportional to $\lambda_B$ violates B, while
the other three terms (which have been obtained from 
eq.~(\ref{wrparity}) by replacing $\widehat H_1$ with $\widehat L$)
violate L.  Phenomenological constraints on
various low-energy B and L violating processes yield limits on each of
the individual coefficients in eq.~(\ref{nrsuppot})
taken one at a time \cite{dreiner}.
If more than one coefficient is simultaneously active, the limits are in
general more complex.  All four terms in eq.~(\ref{nrsuppot})
cannot be simultaneously present and unsuppressed; otherwise 
the proton decay rate would be many orders of magnitude larger than
the present experimental bound.   One way to avoid proton decay is to
impose either B or L separately.  For example, if B is conserved but L
is not, then $\lambda_B=0$ (while the other three terms in
eq.~(\ref{nrsuppot}) can be present).  Such a model 
violates R-parity but preserves a ${\bf Z}_3$ baryon parity.

If R-parity is not conserved, supersymmetric phenomenology exhibits
features that are quite distinct from that of the MSSM.  The LSP is no
longer stable, which implies that not all supersymmetric decay chains
must yield missing energy events at colliders.  Both 
$\Delta L\!=\! 1$ and
$\Delta L\!=\! 2$ phenomena are allowed (assuming B is conserved), 
leading to neutrino masses
and mixing \cite{numass}, neutrinoless double beta decay \cite{0nubb},
sneutrino-antisneutrino mixing \cite{grosshab,otherguys},
and \hbox{$s$-channel} 
resonant production of the sneutrino in $e^+e^-$
collisions \cite{schannel}.   Since the distinction between the Higgs
and matter
multiplets is lost, R-parity violation permits the mixing of sleptons
and Higgs bosons, the mixing of neutrinos and neutralinos, and the
mixing of charged leptons and charginos, leading to more complicated
mass matrices and mass eigenstates.

Squarks can be regarded as
leptoquarks since the term in eq.~(\ref{nrsuppot}) proportional to
$\lambda_L^\prime$ permits processes such as:
\beqa \label{leptoquarks}
&& e^+\overline u_m\to \overline{\widetilde d}_n\to e^+\overline u_m\,,\,
               \overline\nu\overline d_m \nonumber \\
&&e^+ d_m \to \widetilde u_n\to e^+d_m\,.
\eeqa
These processes have received much attention during the past year as a
possible explanation for the HERA high $Q^2$ anomaly \cite{hera}.  Note
that the
same term responsible for the processes displayed above could also generate
purely hadronic decays for sleptons: {\it e.g.}, $\widetilde\ell^-_p\to
\overline u_m d_n$ and $\widetilde\nu_p\to \overline q_m q_n$ ($q=u$ or
$d$).  If such decays were dominant, then the pair production of
sleptons in $e^+e^-$ events would lead to hadronic four-jet
events \cite{fourjet}, a
signature quite different from the missing energy signals expected in
the MSSM.

\section{THE STATUS OF LOW-ENERGY SUPERSYMMETRY}

The organizers of SUSY-97 asked me to summarize the status of
low-energy supersymmetry.

$\bullet$ {\it Theory}. The origin of the soft supersymmetry breaking
terms and the details of their structure remain a mystery.  The
interplay of supersymmetry and the origin of flavor needs elucidation.
There are many ideas but as yet no compelling models.
We do not understand the mechanism that requires nature to lie near one
of the exceptional points in MSSM-124 parameter space.

$\bullet$ {\it Experiment}.  Supersymmetric particles have not yet been
discovered.  Indirect hints such as gauge coupling unification, the
existence of dark matter (for which the LSP is a natural candidate), and
a few interesting collider ``zoo'' events are intriguing but not yet
compelling.

$\bullet$ {\it Phenomenology}.  Canonical supersymmetric signatures at
future colliders are well analyzed and understood.  Much of the recent
efforts have been directed at trying to develop strategies for
precision measurements to prove the underlying supersymmetric structure
of the interactions and to distinguish among models  \cite{snowbagger}.
However, we
are far from understanding all possible facets of MSSM-124 parameter
space (even restricted to those regions that are phenomenologically
viable).  Moreover, the phenomenology of alternative low-energy
supersymmetric models (such as models with R-parity violation)
and its consequences for collider physics have
only recently begun to attract significant attention.
The variety of possible non-minimal models of low-energy supersymmetry
presents an additional challenge to experimenters who plan on searching
for supersymmetry at future colliders.

Low-energy supersymmetry remains the most elegant solution to the
naturalness and hierarchy problems, while
providing a possible link to Planck scale physics and the unification of
particle physics and gravity.  If nature has chosen this path,
then the future of the SUSY-yy workshops is indeed bright.

\section*{ACKNOWLEDGMENTS}
This work was supported in part by the U.S. Department of Energy.
I am grateful to Michael Dine, Stephen Martin and Damien Pierce for
their critical reading of this manuscript.

\end{document}